\documentclass[journal]{IEEEtran}

\ifCLASSINFOpdf
\else
   \usepackage[dvips]{graphicx}
\fi
\usepackage{url}

\ifCLASSOPTIONcompsoc
\usepackage[caption=false,font=normalsize,labelfont=sf,textfont=sf]{subfig}
\else
\usepackage[caption=false,font=footnotesize]{subfig}
\fi

\usepackage{graphicx}
\usepackage{amsmath}
\usepackage{amssymb}
\usepackage{amsfonts}
\usepackage{multirow}
\usepackage{algpseudocode}
\usepackage[utf8]{inputenc}
\usepackage[T1]{fontenc}
\usepackage[absolute,overlay,showboxes]{textpos}

\newcommand{\M}[1]{\mathbf{#1}}
\newcommand{\BS}[1]{\boldsymbol{#1}}

\newcommand{\realm}{\mbox{REAL-M}}

\begin{document}

\TPMargin{1mm}
\begin{textblock*}{17cm}(2.3cm,26.6cm)
\scriptsize
\noindent
\copyright \ 2022 IEEE.  Personal use of this material is permitted.  Permission from IEEE must be obtained for all other uses, in any current or future media, including reprinting/republishing this material for advertising or promotional purposes, creating new collective works, for resale or redistribution to servers or lists, or reuse of any copyrighted component of this work in other works.
This article has been accepted for publication in IEEE Signal Processing Letters. This is the author's version which has not been fully edited and content may change prior to final publication. Citation information: DOI 10.1109/LSP.2022.3232276
\end{textblock*}

\title{MixCycle: Unsupervised Speech Separation via Cyclic Mixture Permutation Invariant Training}

\author{Ertuğ Karamatlı and Serap Kırbız
\thanks{Submitted for review on November 5, 2022.}
\thanks{E. Karamatlı is with the Department of Computer Engineering, Boğaziçi University, Istanbul, Turkey (e-mail: ertug@karamatli.com).}
\thanks{S. Kırbız is with the Department of Electrical and Electronics Engineering, MEF University, Istanbul, Turkey (e-mail: kirbizs@mef.edu.tr).}}

\markboth{Submitted to the IEEE Signal Processing Letters}
{Shell \MakeLowercase{\textit{et al.}}: Bare Demo of IEEEtran.cls for IEEE Journals}
\maketitle

\begin{abstract}
We introduce two unsupervised source separation methods, which involve self-supervised training from single-channel two-source speech mixtures. Our first method, mixture permutation invariant training (MixPIT), enables learning a neural network model which separates the underlying sources via a challenging proxy task without supervision from the reference sources. Our second method, cyclic mixture permutation invariant training (MixCycle), uses MixPIT as a building block in a cyclic fashion for continuous learning. MixCycle gradually converts the problem from separating mixtures of mixtures into separating single mixtures. We compare our methods to common supervised and unsupervised baselines: permutation invariant training with dynamic mixing (PIT-DM) and mixture invariant training (MixIT). We show that MixCycle outperforms MixIT and reaches a performance level very close to the supervised baseline (PIT-DM) while circumventing the over-separation issue of MixIT. Also, we propose a self-evaluation technique inspired by MixCycle that estimates model performance without utilizing any reference sources. We show that it yields results consistent with an evaluation on reference sources (LibriMix) and also with an informal listening test conducted on a real-life mixtures dataset (\realm{}).
\end{abstract}


\IEEEpeerreviewmaketitle

\vspace{-0.25cm}
\section{Introduction}
\IEEEPARstart{R}{ecent} state-of-the-art speech separation methods \cite{subakan2021attention,nachmani2020voice,tzinis2020sudo}, which employ permutation invariant training (PIT) \cite{yu2017permutation,kolbaek2017multitalker}, have achieved separation results with almost no perceptible distortion. However, supervised speech separation requires a large dataset of mixture recordings and the corresponding ground truth source recordings, which is challenging and impractical to acquire in the same acoustic environment \cite{subakan2022real}. Therefore, current methods are usually trained on synthetic mixtures that are generated by mixing clean single-speaker recordings which may not reflect real-life mixture recordings and still require a large dataset of clean recordings. 

To avoid this data collection problem, weakly-supervised methods \cite{karamatli2019audio,pishdadian2020finding}, unsupervised methods \cite{drude2019unsupervised,bando2021neural,neri2021unsupervised} and self-supervised representation learning \cite{huang2022investigating} can be employed. Recently, mixture invariant training (MixIT) \cite{wisdom2020unsupervised} has been proposed which 
enables unsupervised training by using single-channel mixtures as references and artificial mixture of mixtures (MoMs) as input. However, MixIT estimates a greater number of sources than the number of underlying sources in the test stage which can cause an over-separation problem where parts of the source signals get spread out between the outputs. In \cite{wisdom2021sparse}, applying sparsity, covariance and classification losses to MixIT is proposed to lessen the over-separation issue, and also a computationally efficient approximation is introduced to handle a larger number of sources.
In \cite{sivaraman2022adapting}, adapting speech separation to real-world meetings using MixIT is proposed. Teacher-student MixIT \cite{zhang2021teacher} addresses the over-separation issue of MixIT by training another model where the number of outputs matches the number of underlying sources.

Mixup-Breakdown \cite{lam2020mixup} is a semi-supervised separation method with a mean-teacher \cite{tarvainen2017mean} model that improves the generalization capability to mismatch conditions. Although our teacher-student arrangement is similar, our methods are trained from scratch in a purely unsupervised manner and do not require the calculation of moving averages to update the teacher parameters (i.e. simpler implementation, less resource demand). Also, we remix pairs of the teacher source estimates originating from different mixtures instead of remixing pairs from the same mixtures so that our student model is trained on an extremely large number of unique mixtures.

RemixIT \cite{tzinis2022continual} is a self-supervised method for speech enhancement. Although our remixing strategy is reminiscent of the bootstrapped remixing approach in \cite{tzinis2022continual}, our methods are purely unsupervised and do not require supervised pre-training. Also, our methods avoid the over-separation problem by using PIT instead of MixIT.

In this work, we explore training source separation models without having access to the ground truth source signals that constitute single-channel two-source speech mixtures. Our main contributions are summarized as follows:
\begin{enumerate}

\item We present two purely unsupervised source separation methods that are based on self-supervised \cite{weng2019selfsup} training: MixPIT and MixCycle. MixPIT uses a challenging proxy task to avoid the over-separation problem of MixIT \cite{wisdom2020unsupervised}. MixCycle uses MixPIT as a building block and applies it in a cyclic fashion for continuous learning.

\item On a standard speech separation dataset (LibriMix) \cite{cosentino2020librimix}, we observe that MixCycle attains a performance close to supervised training, which is uncommon in the related works. We also observe that training it on only 5\% of the dataset obtains a performance close to training on 100\% of the dataset, which demonstrates its data efficiency.
\item We propose a self-evaluation technique inspired by MixCycle, which estimates scale-invariant signal-to-noise ratio improvement (SI-SNRi) \cite{le2019sdr} without any access to the reference sources. It yields similar results to a ground-truth evaluation on LibriMix and consistent results with an informal listening test we conducted on a real-life mixtures dataset (\realm{} \cite{subakan2022real}).
\end{enumerate}

\section{Background}
\label{sec:background}

\vspace{-0.1cm}
\subsection{Permutation Invariant Training (PIT)}
\label{sec:pit}
\begin{figure}
\centerline{\includegraphics[width=0.65\columnwidth]{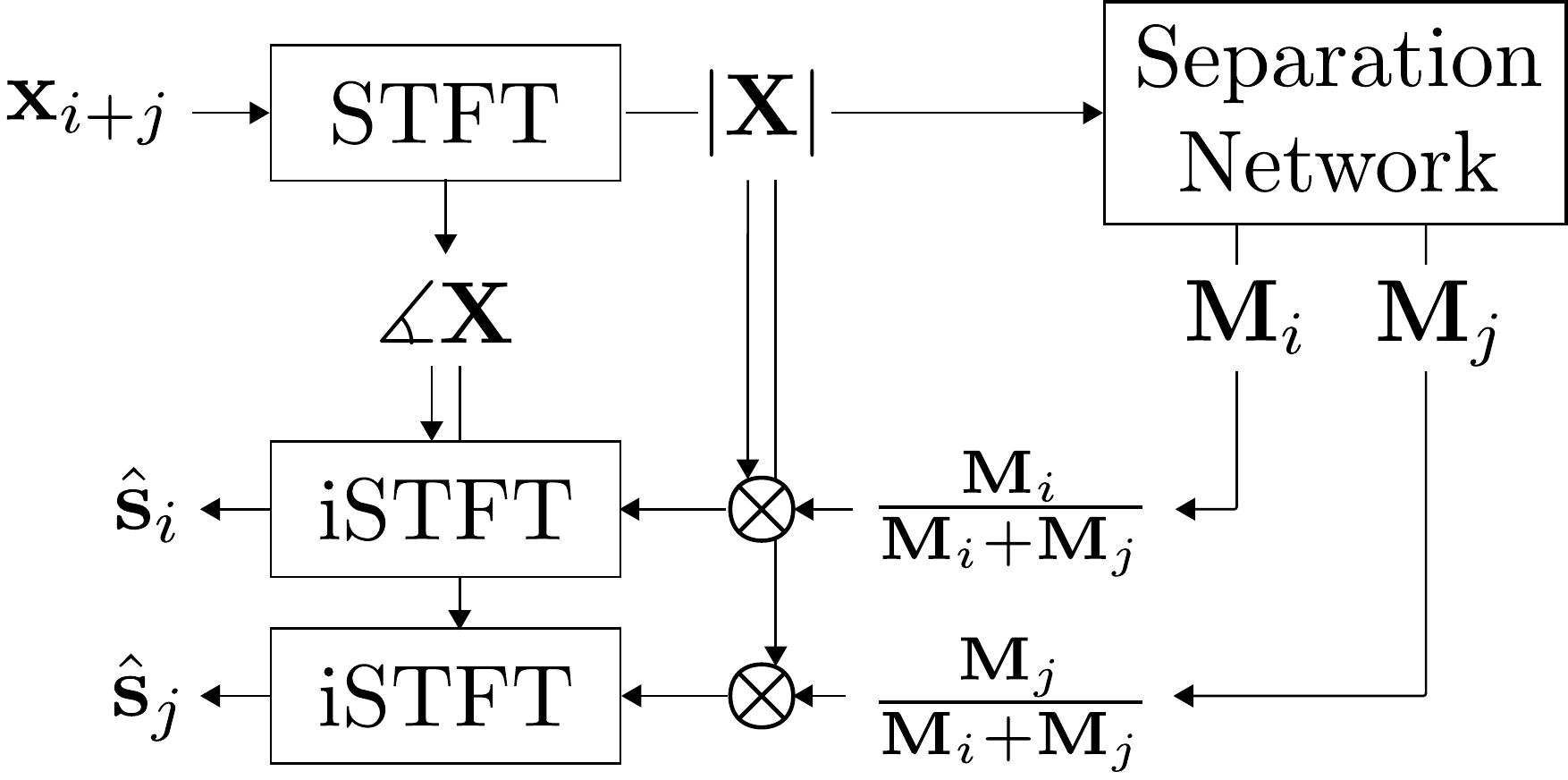}}
\caption{Our base model $f_{\boldsymbol{\theta}}$ with parameters $\BS{\theta}$. The time-domain mixture signal $\M{x}_{i+j}$ is converted into magnitude $\M{\left|X\right|}$ and phase $\measuredangle{\M{X}}$ via short-time Fourier transform (STFT). The separation network is based on the Conv-TasNet \protect\cite{luo2019conv} architecture, but any other architecture can also be used. The separation network is supplied with the mixture magnitude $\M{\left|X\right|}$, and it estimates the masks $\M{M}_i$ and $\M{M}_j$. The masks are ensured to add up to one and multiplied with the mixture magnitude $\M{\left|X\right|}$. The resulting source magnitude estimates are passed through a differentiable inverse STFT (iSTFT) to arrive at the time-domain source signal estimates $\hat{\M{s}}_i$ and $\hat{\M{s}}_j$.}
\label{fig:base_model}
\vspace{-0.4cm}
\end{figure}

We define a supervised training dataset $\mathcal{X}_s=\{(\M{x}_{i+j}, \M{s}_i, \M{s}_j)\}_{i+j}$ where $\M{x}_{i+j} = \M{s}_i + \M{s}_j$ is a mixture signal of the time-domain source signals $\M{s}_i, \M{s}_j \in \mathbb{R}^L$ with length $L$.
The model $\hat{\M{S}}=f_{\BS{\theta}}(\M{x}_{i+j})$ outputs the source estimates $\hat{\M{s}}_i$, $\hat{\M{s}}_j$ in the rows of $\hat{\M{S}}$. The details of our base model $f_{\BS{\theta}}$ are given in Fig.~\ref{fig:base_model}. The loss function for utterance-level PIT \cite{kolbaek2017multitalker} is
\begin{equation}
\mathcal{L}_{\mathrm{PIT}}(\M{s}_i,\M{s}_j,\hat{\M{S}}) = \min_{\M{P}} \left[ \mathcal{L}(\M{s}_i, [\M{P}\hat{\M{S}}]_1) + \mathcal{L}(\M{s}_j, [\M{P}\hat{\M{S}}]_2) \right],
\label{eq:pit}
\end{equation}
where $\M{P}$ is a $2{\times}2$ permutation matrix, $[\cdot]_r$ selects the $r$-th row of a matrix and $\mathcal{L}$ is the loss function calculated between reference sources and their estimates, as illustrated in Fig.~\ref{fig:models_pit}.

\vspace{-0.4cm}
\subsection{Mixture Invariant Training (MixIT)}
\begin{figure*}[!t]
\centering
\subfloat[PIT]{\includegraphics[height=1.25in]{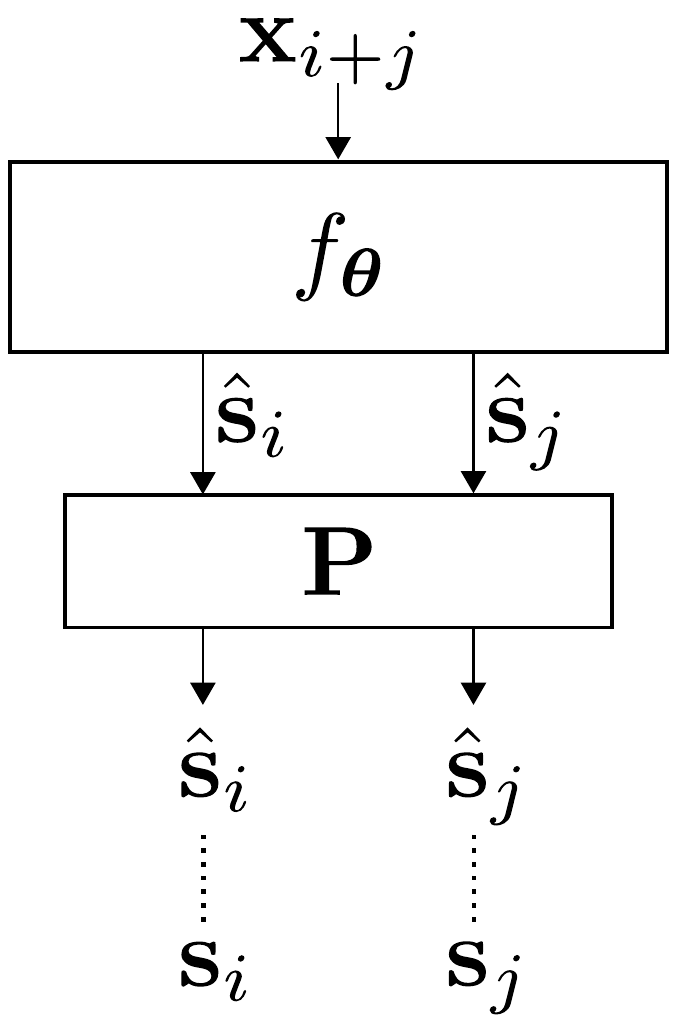}
\label{fig:models_pit}}
\hfil
\subfloat[MixIT]{\includegraphics[height=1.25in]{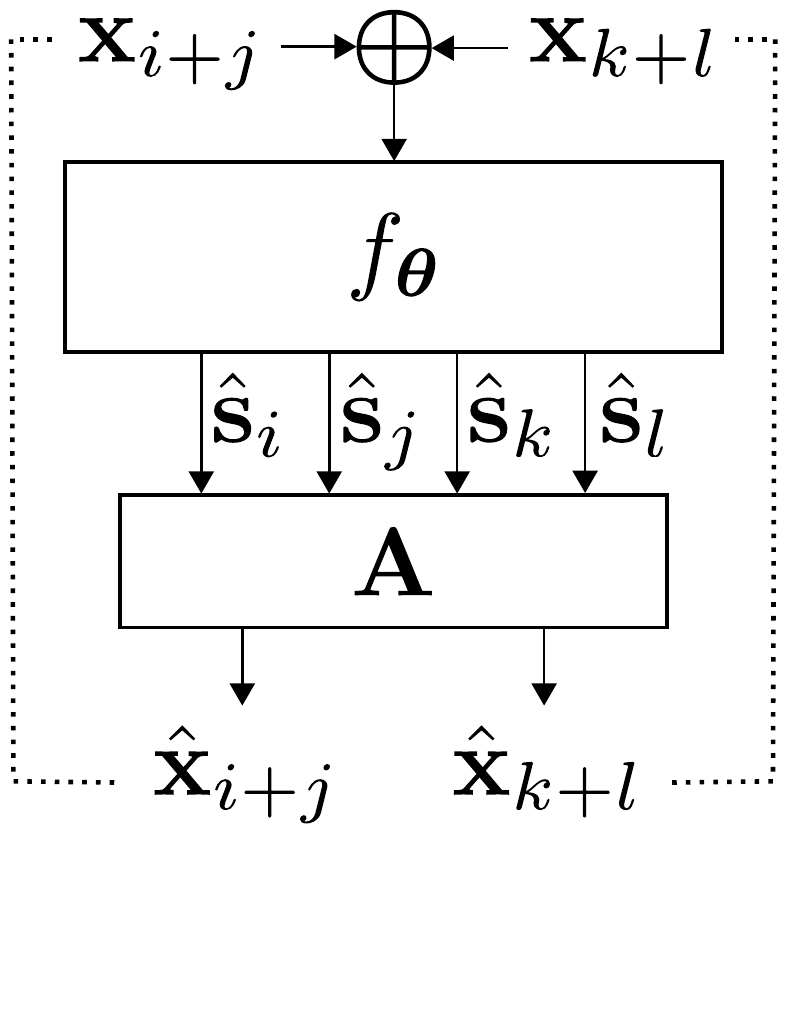}
\label{fig:models_mixit}}
\hfil
\subfloat[MixPIT (proposed)]{\includegraphics[height=1.25in]{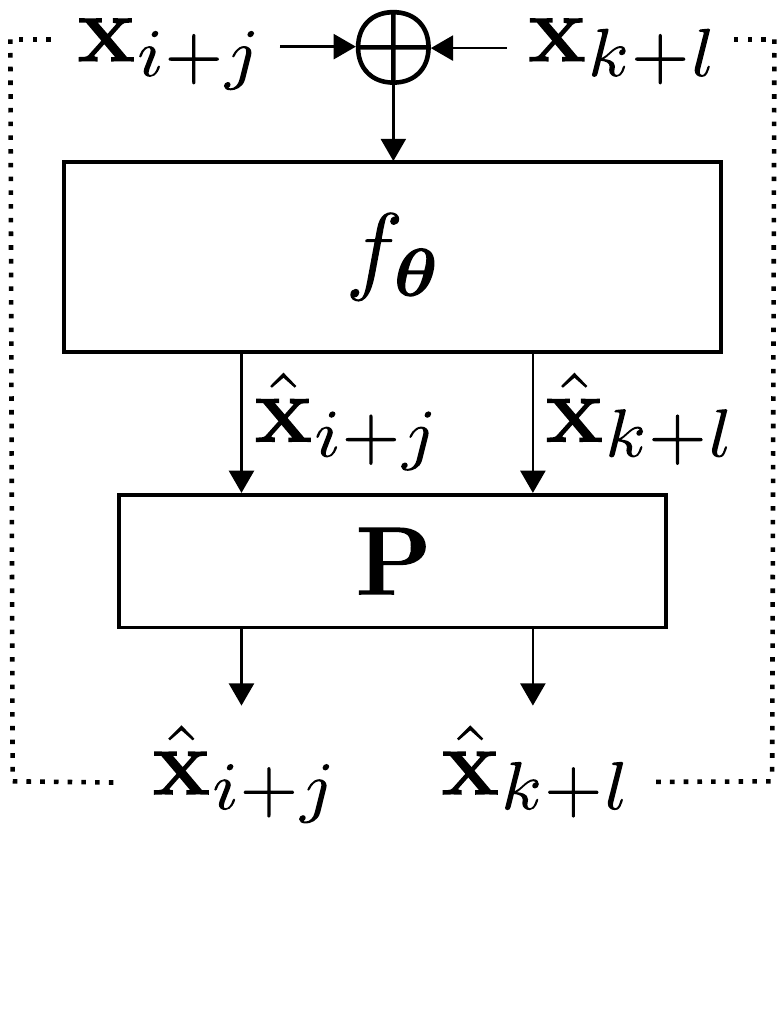}
\label{fig:models_mixpit}}
\hfil
\subfloat[MixCycle (proposed)]{\includegraphics[height=1.25in]{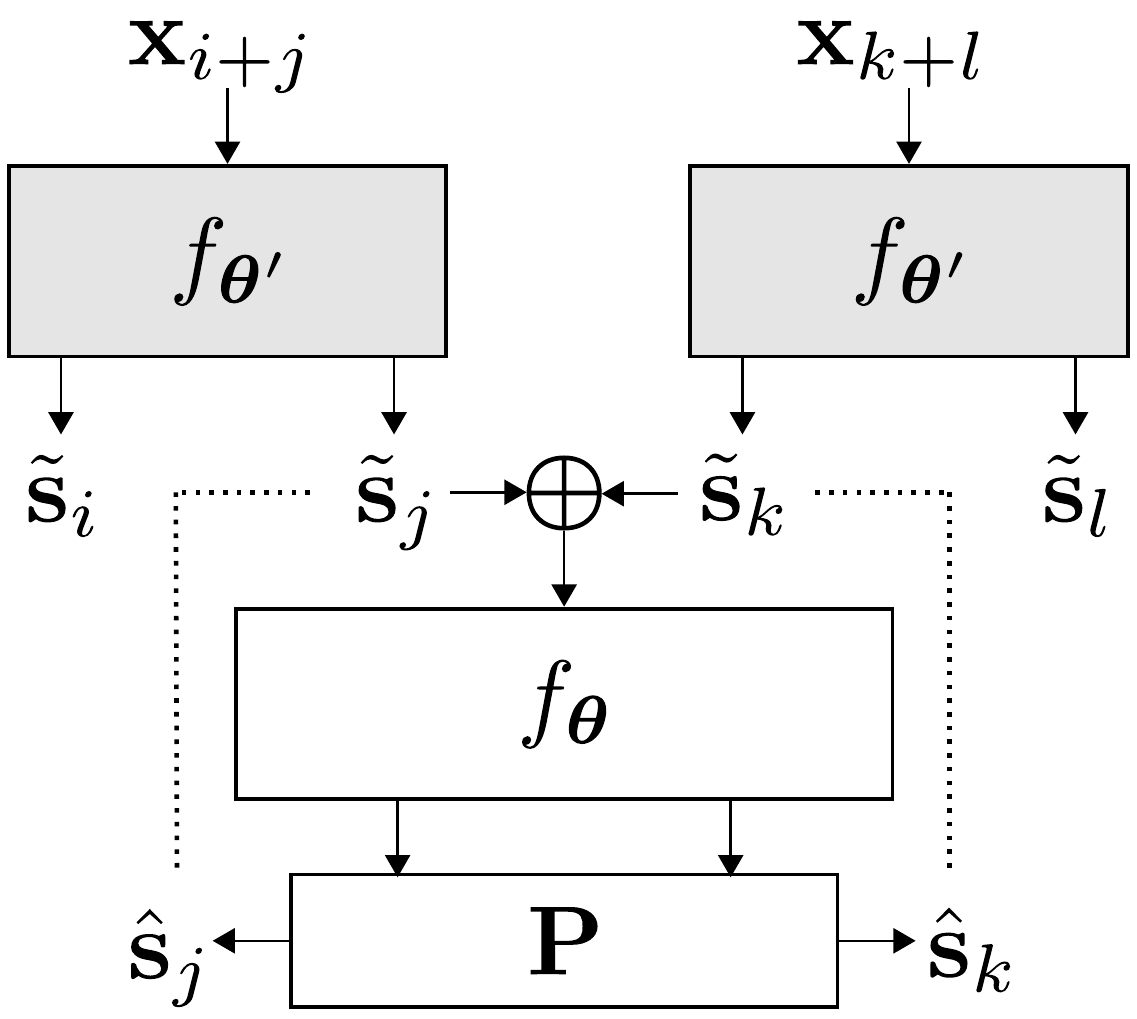}
\label{fig:models_mixcycle}}
\hfil
\subfloat[Self-evaluation (proposed)]{\includegraphics[height=1.25in]{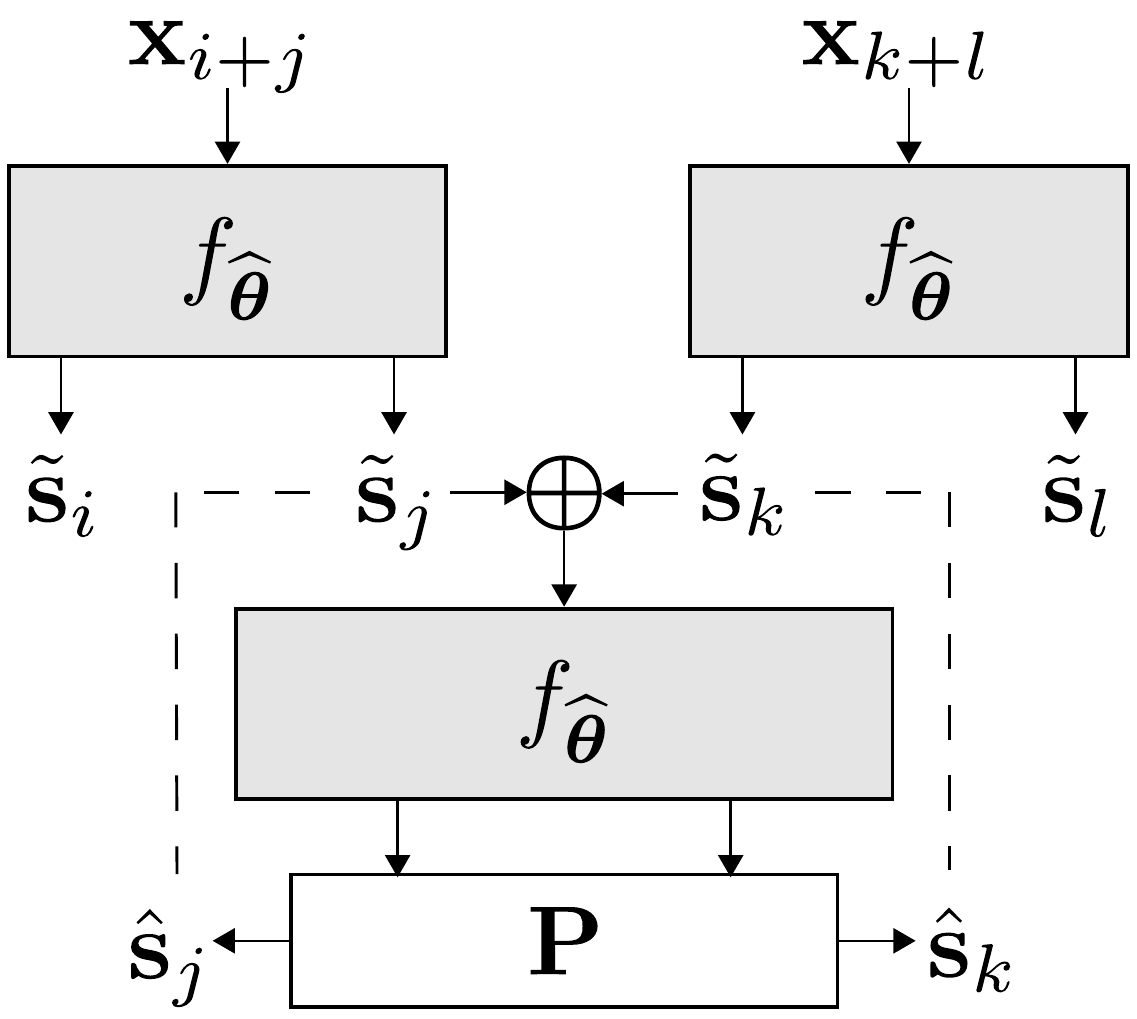}
\label{fig:models_selfeval}}
\caption{Illustration of the methods that we consider in this work. The dotted lines denote the calculation of the loss function $\mathcal{L}$ between various signals. \protect\subref{fig:models_pit}~In PIT, the mixture $\M{x}_{i+j}$ is inputted to the model $f_{\boldsymbol{\theta}}$. The best match between the two model outputs and the sources $\M{s}_i$, $\M{s}_j$ is chosen by a permutation matrix $\M{P}$ to update the parameters $\boldsymbol{\theta}$. \protect\subref{fig:models_mixit}~In MixIT, the mixtures $\M{x}_{i+j}$, $\M{x}_{k+l}$ are summed and inputted to the model $f_{\boldsymbol{\theta}}$. The best match between the four model outputs and the mixtures $\M{x}_{i+j}$, $\M{x}_{k+l}$ is chosen by a mixing matrix $\M{A}$ to update the parameters $\boldsymbol{\theta}$. \protect\subref{fig:models_mixpit}~In MixPIT, the mixtures $\M{x}_{i+j}$, $\M{x}_{k+l}$ are summed and inputted to the model $f_{\boldsymbol{\theta}}$. The best match between the two model outputs and the mixtures $\M{x}_{i+j}$, $\M{x}_{k+l}$ is chosen by a permutation matrix $\M{P}$ to update the parameters $\boldsymbol{\theta}$. \protect\subref{fig:models_mixcycle}~In MixCycle, the mixtures $\M{x}_{i+j}$, $\M{x}_{k+l}$ are inputted to the teacher model $f_{\boldsymbol{\theta}'}$. The teacher outputs are paired such that each pair contains source estimates from two different original mixtures: $\{\tilde{\M{s}}_j,\tilde{\M{s}}_k\}$, $\{\tilde{\M{s}}_i, \tilde{\M{s}}_l\}$. Each pair is summed to produce an artificial mixture before being inputted to the student model $f_{\boldsymbol{\theta}}$. Considering the pair $\{\tilde{\M{s}}_j,\tilde{\M{s}}_k\}$, the best match between the two student outputs and the teacher source estimates $\tilde{\M{s}}_j$, $\tilde{\M{s}}_k$ is chosen by a permutation matrix $\M{P}$ to update the parameters $\boldsymbol{\theta}$ while $\boldsymbol{\theta}'$ are frozen. The other pair $\{\tilde{\M{s}}_i, \tilde{\M{s}}_l\}$ is also processed similarly. \protect\subref{fig:models_selfeval}~In self-evaluation, given a trained model $f_{\widehat{\BS{\theta}}}$, the ground-truth reference sources are estimated as $\tilde{\M{s}}_i,\tilde{\M{s}}_j,\tilde{\M{s}}_k,\tilde{\M{s}}_l$. These estimated reference sources are paired and summed as in MixCycle. The trained model $f_{\widehat{\BS{\theta}}}$ is evaluated on the resulting artificial mixtures where the dashed lines denote the calculation of SI-SNRi between the source estimates $\hat{\M{s}}_j,\hat{\M{s}}_k$ and the estimated reference sources $\tilde{\M{s}}_j,\tilde{\M{s}}_k$.}
\label{fig_sim}
\vspace{-0.25cm}
\end{figure*}

We define an unsupervised training dataset $\mathcal{X}_u=\{(\M{x}_{i+j}, \M{x}_{k+l})\}_{(i+j,k+l)}$. The input to the model is formed by summing two mixtures $\M{x}_{i+j}$ and $\M{x}_{k+l}$. The model $\hat{\M{S}}=f_{\BS{\theta}}(\M{x}_{i+j}+\M{x}_{k+l})$ outputs the source estimates $\hat{\M{s}}_i$, $\hat{\M{s}}_j$, $\hat{\M{s}}_k$, $\hat{\M{s}}_l$. The loss function for MixIT \cite{wisdom2020unsupervised} is calculated between reference mixtures and their estimates as
\begin{equation}
\mathcal{L}_{\mathrm{MixIT}}(\cdot) = \min_{\M{A}} \left[ \mathcal{L}(\M{x}_{i+j}, [\M{A}\hat{\M{S}}]_1) + \mathcal{L}(\M{x}_{k+l}, [\M{A}\hat{\M{S}}]_2) \right]
\end{equation}
where $\M{A}$ is a $2{\times}4$ binary mixing matrix, as illustrated in Fig.~\ref{fig:models_mixit}. In the test stage, a single mixture is supplied to the model $f_{\BS{\theta}}(\M{x}_{i+j})$.

\vspace{-0.25cm}
\section{Proposed Methods}
\label{sec:model}

\subsection{Mixture Permutation Invariant Training (MixPIT)}
The main limitation of MixIT is the over-separation problem which stems from having a greater number of model outputs than the actual number of underlying sources. Here, we remove this limitation by training a model where we have four sources (i.e. a mixture of mixtures) to separate but only have two model outputs. We use the model $f_{\BS{\theta}}(\M{x}_{i+j}+\M{x}_{k+l})$ which has three possible output pairs $\hat{\M{S}}_1,\hat{\M{S}}_2,\hat{\M{S}}_3$ as given in Table~\ref{table:mixpit_outputs}, ignoring the output permutations (abbreviated to "Perm."). The case of the output pair $\hat{\M{S}}_1$ is illustrated in Fig.~\ref{fig:models_mixpit}.

Assuming that the sources $\M{s}_i, \M{s}_j, \M{s}_k, \M{s}_l$ are statistically independent from each other, the output pairs $\hat{\M{S}}_1,\hat{\M{S}}_2,\hat{\M{S}}_3$ are equally likely because the model cannot learn the difference between the possible source pairings of the input mixtures: $\M{x}_{i+j}+\M{x}_{k+l}$, $ \M{x}_{i+k}+\M{x}_{j+l}$ and $\M{x}_{j+k}+\M{x}_{i+l}$.

We train the model $f_{\BS{\theta}}$ using the loss function
\begin{equation}
\mathcal{L}_{\mathrm{MixPIT}}(\cdot)=\mathcal{L}_{\mathrm{PIT}}(\M{x}_{i+j},\M{x}_{k+l}, \hat{\M{S}}),
\end{equation}
where $\hat{\M{S}} \in \{\hat{\M{S}}_1,\hat{\M{S}}_2,\hat{\M{S}}_3\}$ is the model output pair. Table~\ref{table:mixpit_outputs} lists all of the possible instances of $\mathcal{L}_{\mathrm{MixPIT}}(\cdot)$. For each of these instances, matching source indices (MSI) are listed as well.

The PIT loss $\mathcal{L}_{\mathrm{PIT}}(\cdot)$ in (\ref{eq:pit}) selects the best output permutation that minimizes the loss value. Therefore, the output pair $\hat{\M{S}}_1$ corresponds to an \emph{exact match} where all of the sources match (Perm. 1 is always selected due to having the lower loss value), while $\hat{\M{S}}_2$ and $\hat{\M{S}}_3$ correspond to a \emph{partial match} where two of the sources match (Perm. 1 and 2 are equally likely). Due to the fact that at least two sources are guaranteed to match in the loss function, the model learns to separate two sources as well, despite some noise from mismatching sources.

Ultimately, the model is trained with the challenging proxy task of separating mixtures of mixtures which also covers our main objective of separating single mixtures. In the test stage, we supply the model with a single mixture $f_{\BS{\theta}}(\M{x}_{i+j})$ to obtain the source estimates $\hat{\M{s}}_{i}$ and $\hat{\M{s}}_{j}$. There is no over-separation due to the equal number of model outputs and underlying sources.

\begin{table}
\caption{All of the possible cases for the MixPIT loss function}
\label{table:mixpit_outputs}
\setlength{\tabcolsep}{3pt}
\centering
\begin{tabular}{|l|c|l|r|}
\hline
\textbf{Model output pair} & \textbf{Perm.} & \textbf{Loss function} $\mathcal{L}_{\mathrm{MixPIT}}(\cdot)$ & \textbf{MSI} \\
\hline
\multirow{2}{*}{$\hat{\M{S}}_1=\{\hat{\M{x}}_{i+j},\hat{\M{x}}_{k+l}\}$} & 1 & $\mathcal{L}(\M{x}_{i+j},\hat{\M{x}}_{i+j})+\mathcal{L}(\M{x}_{k+l},\hat{\M{x}}_{k+l})$ & $i, j, k, l$ \\
 & 2 & $\mathcal{L}(\M{x}_{i+j},\hat{\M{x}}_{k+l})+\mathcal{L}(\M{x}_{k+l},\hat{\M{x}}_{i+j})$ & -- \\
\hline
\multirow{2}{*}{$\hat{\M{S}}_2=\{\hat{\M{x}}_{i+k},\hat{\M{x}}_{j+l}\}$} & 1 & $\mathcal{L}(\M{x}_{i+j},\hat{\M{x}}_{i+k})+\mathcal{L}(\M{x}_{k+l},\hat{\M{x}}_{j+l})$ & $i, l$ \\
 & 2 & $\mathcal{L}(\M{x}_{i+j},\hat{\M{x}}_{j+l})+\mathcal{L}(\M{x}_{k+l},\hat{\M{x}}_{i+k})$ & $j, k$ \\
\hline
\multirow{2}{*}{$\hat{\M{S}}_3=\{\hat{\M{x}}_{j+k},\hat{\M{x}}_{i+l}\}$} & 1 & $\mathcal{L}(\M{x}_{i+j},\hat{\M{x}}_{j+k})+\mathcal{L}(\M{x}_{k+l},\hat{\M{x}}_{i+l})$ & $j, l$ \\
 & 2 & $\mathcal{L}(\M{x}_{i+j},\hat{\M{x}}_{i+l})+\mathcal{L}(\M{x}_{k+l},\hat{\M{x}}_{j+k})$ & $i, k$ \\
\hline
\end{tabular}
\vspace{-0.2cm}
\end{table}

\vspace{-0.1cm}
\subsection{Cyclic Mixture Permutation Invariant Training (MixCycle)}
Here we propose a new method that improves the performance further by using MixPIT as a building block and applying it in a cyclic fashion for continuous learning.

First, we use a teacher model $f_{\BS{\theta}'}$ to estimate four sources from the two input mixtures
\begin{align}
&\{\tilde{\M{s}}_i,\tilde{\M{s}}_j\}=f_{\BS{\theta}'}(\M{x}_{i+j}),\quad  \{\tilde{\M{s}}_k,\tilde{\M{s}}_l\}=f_{\BS{\theta}'}(\M{x}_{k+l})
\label{eq:teacher}
\end{align}
where $\BS{\theta}'=\BS{\theta}^{(\tau-1)}$ are the parameters at the previous training step $\tau-1$. Second, we use these estimated sources to generate unique mixtures such that each constituent source estimate originates from a different randomly-selected mixture. We accomplish this by randomly choosing one of the following two remixing options, which eliminates any bias due to model output permutations:
\begin{align}
\text{Opt. 1:} \quad&\tilde{\M{x}}_{i+k}=\tilde{\M{s}}_i+\tilde{\M{s}}_k, \quad \tilde{\M{x}}_{j+l}=\tilde{\M{s}}_j+\tilde{\M{s}}_l \\
\text{Opt. 2:} \quad&\tilde{\M{x}}_{j+k}=\tilde{\M{s}}_j+\tilde{\M{s}}_k, \quad \tilde{\M{x}}_{i+l}=\tilde{\M{s}}_i+\tilde{\M{s}}_l.
\label{eq:genmix}
\end{align}
Finally, we train a student model $f_{\BS{\theta}}$ on these artificial mixtures with the loss function (no backpropagation into $f_{\BS{\theta}'}$)
\begin{equation}
    \mathcal{L}_{\mathrm{MixCycle}}(\cdot)=\mathcal{L}_{\mathrm{PIT}}(\tilde{\M{s}}_j,\tilde{\M{s}}_k, \hat{\M{S}}_1)+\mathcal{L}_{\mathrm{PIT}}(\tilde{\M{s}}_i,\tilde{\M{s}}_l, \hat{\M{S}}_2)
    \label{eq:MixCycleloss}
\end{equation}
to estimate the sources (assuming Opt. 2 is chosen):
\begin{gather}
\hat{\M{S}}_1=\{\hat{\M{s}}_j,\hat{\M{s}}_k\}=f_{\BS{\theta}}(\tilde{\M{x}}_{j+k}),\quad  \hat{\M{S}}_2=\{\hat{\M{s}}_i,\hat{\M{s}}_l\}=f_{\BS{\theta}}(\tilde{\M{x}}_{i+l}) 
\label{eq:student}
\end{gather}
where $\BS{\theta}=\BS{\theta}^{(\tau)}$ are the parameters at the current training step $\tau$.
The model is illustrated in Fig.~\ref{fig:models_mixcycle}.

We designed the model $f_{\BS{\theta}}$ such that it produces informative initial source estimates and helps prevent the source estimates from diverging throughout the training process. We accomplish this by employing time-frequency masking and ensuring that the masks add up to one as given in Fig.~\ref{fig:base_model}. Therefore, we have mixture consistency \cite{wisdom2019differentiable} as $\tilde{\M{s}}_i+\tilde{\M{s}}_j=\M{x}_{i+j}$. Also, our remixing strategy acts as a data augmentation mechanism by generating an extremely large number of unique mixtures on-the-fly. Consequently, it increases the effective size of the available training set, similar to dynamic mixing \cite{zeghidour2021wavesplit}.

MixCycle can be viewed as a cascade of successive MixPIT training steps with a continuously improved mixture input instead of a static mixture of mixtures input. To elaborate on this, we define the initial source estimates $\tilde{\M{s}}_j^{(0)},\tilde{\M{s}}_k^{(0)}$ of the teacher model $f_{\BS{\theta}'}$ in (\ref{eq:teacher}) when the parameters $\BS{\theta}'=\BS{\theta}^{(0)}$ are randomly initialized at the first training step $\tau=1$:
\begin{align}
\label{eq:init_source_1}
\tilde{\M{s}}_j^{(0)} &= \operatorname{iSTFT}\left(|\M{X}_{i+j}|\frac{\M{M}_j^{(0)}}{\M{M}_i^{(0)}+\M{M}_j^{(0)}},\measuredangle{\M{X}_{i+j}}\right)\\
\label{eq:init_source_2}
\tilde{\M{s}}_k^{(0)} &= \operatorname{iSTFT}\left(|\M{X}_{k+l}|\frac{\M{M}_k^{(0)}}{\M{M}_k^{(0)}+\M{M}_l^{(0)}},\measuredangle{\M{X}_{k+l}}\right)
\end{align}
where $|\M{X}_{i+j}|,|\M{X}_{k+l}|$ and $\measuredangle{\M{X}_{i+j}},\measuredangle{\M{X}_{k+l}}$ are the magnitude and phase spectrograms of the mixture signals $\M{x}_{i+j},\M{x}_{k+l}$, respectively.
$\M{M}_i^{(0)},\M{M}_j^{(0)},\M{M}_k^{(0)},\M{M}_l^{(0)} \in (0,1)^{F\times T}$ are noisy mask outputs of the randomly initialized model with $F$ frequency bins and $T$ time frames. 
If we consider the initial input mixture $\tilde{\M{x}}_{j+k}^{(0)}=\tilde{\M{s}}_j^{(0)}+\tilde{\M{s}}_k^{(0)}$ of the student model $f_{\BS{\theta}}(\cdot)$ in (\ref{eq:student}) using (\ref{eq:init_source_1}) and (\ref{eq:init_source_2}), we can see that this is similar to the proposed MixPIT method because we have a mixture of noisy mixtures $\tilde{\M{x}}_{j+k}^{(0)}$ as input and try to separate it into single noisy mixtures $\tilde{\M{s}}_j^{(0)}$, $\tilde{\M{s}}_k^{(0)}$. In contrast to MixPIT, the input mixture $\tilde{\M{x}}_{j+k}$ is not static and refined at each training step $\tau$ such that the constituent teacher source estimates $\tilde{\M{s}}_j$, $\tilde{\M{s}}_k$, which start as noisy copies of the original mixtures, are transformed into accurate estimates of the corresponding sources $\M{s}_j$, $\M{s}_k$ as we optimize the parameters $\BS{\theta}$.

In practice, the initial source estimates are very noisy due to the random initialization. To reduce the noise and stabilize the training process, we initialize the model by training it with the proposed MixPIT method for the first $I$ epochs.

\vspace{-0.1cm}
\section{Experiments}
\label{sec:experiments}

We evaluate the proposed methods on two two-speaker datasets: a standard speech separation dataset (LibriMix \cite{cosentino2020librimix}) and a recently-released real-life mixtures dataset (\realm{} \cite{subakan2022real}) which has no ground-truth reference sources. For LibriMix, we used the clean version of the \texttt{train-360} split with its \emph{min} mode and an 8 kHz sampling rate. The training, validation and test sets contain 212, 11 and 11 hours of speech mixtures, respectively. We refer to the complete training set as the \emph{100\% dataset} and its 5\% random subset as the \emph{5\% dataset} while keeping the original validation and test sets.
For \realm{}, we discarded the "early collection" subset, which showed higher variation in difficulty, and split the remaining data into training and validation sets with 61 and 13 minutes of mixed speech, respectively.

As given in \cite{luo2019conv}, the best performing Conv-TasNet configuration uses $D=8$ dilated convolutions in each repeated block. It also uses a window size of $16$ and a hop size of $8$ for its learned representation. On the other hand, we use STFT/iSTFT with a window size of $512$, a hop size of $128$, and a Hann window. To compensate for our shorter representation, we used $D=4$ which keeps the receptive field of the stacked dilated convolutions similar between the representations.

We randomly sampled three-second-long segments from utterances while training. We used the negative thresholded signal-to-noise ratio with $\mathrm{SNR}_\mathrm{max}=30$~dB as in \cite{wisdom2020unsupervised} for the loss function $\mathcal{L}$. We used the Adam optimization algorithm with its default parameters \cite{Kingma2015} and a batch size of $128$. We applied gradient clipping with a maximum $L_2$-norm of $5$, which was calculated over all gradients together. We employed early stopping and used the models with the best validation scores for testing. We initialized the MixCycle models with MixPIT for $50$ and $250$ epochs on the 100\% and 5\% datasets, respectively. We used LibriMix to initialize the MixCycle model for \realm{} because the \realm{} dataset was too small (${\sim}1$ hour of speech mixtures) for this task.

We evaluate performance using SI-SNRi \cite{le2019sdr}. We find the best match between the reference sources and the model outputs to calculate the SI-SNRi. 
We used PyTorch with an NVIDIA GTX 1080 Ti GPU to develop and evaluate our methods. We released the source code\footnote[1]{\url{https://github.com/ertug/MixCycle}} for reproducibility and published audio samples\footnotemark[1] to demonstrate our results.

\vspace{-0.4cm}
\subsection{Results on LibriMix}
The proposed MixPIT and MixCycle methods are compared with supervised and unsupervised baselines: Conv-TasNet \cite{luo2019conv}, ideal ratio mask (IRM) \cite{wang2014training}, PIT \cite{kolbaek2017multitalker}, PIT with dynamic mixing \cite{zeghidour2021wavesplit} (PIT-DM), and unsupervised MixIT \cite{wisdom2020unsupervised}. Table~\ref{table:results} shows the performance of the methods trained on the 5\% and 100\% datasets (LibriMix) in terms of SI-SNRi, reporting means and standard deviations. We also report the time per training step (Step) in seconds (s) and total training time (Total) in hours (h) on the 100\% dataset. The supervised methods (Super.) can be considered as an empirical upper bound on the performance of the unsupervised methods.
All of the baseline methods are our implementation except Conv-TasNet. 
We observed that the performance of MixPIT is only slightly behind MixIT, despite its challenging training objective and lower computational demand.
We also include an oracle evaluation of the same MixIT model by using the mixing matrix $\M{A}$ to find the best match between the remixed outputs and reference sources as in \cite{zhang2021teacher}. The gap between the oracle and standard evaluation demonstrates the significance of the over-separation issue.

MixCycle improved upon MixPIT as expected and achieved the best performance among the unsupervised methods. Also, it reached a performance level that is close to supervised training (PIT-DM). This is due to the teacher model estimating the sources more accurately as the training progresses; therefore, the student model gets trained on almost the same supervised dataset with dynamic mixing that PIT-DM has access to. We can see the data efficiency of MixCycle on the 5\% dataset as the performance approached that on the 100\% dataset.

\vspace{-0.4cm}
\subsection{Results on \realm{}}
The results on LibriMix in the previous section has shown that a teacher model combined with our remixing strategy produces such accurate artificial mixtures that there is not much difference between training on the artificial (i.e. MixCycle) and original (i.e. PIT-DM) mixtures. Therefore, we propose repurposing MixCycle as a self-evaluation technique to estimate SI-SNRi as illustrated in Fig.~\ref{fig:models_selfeval}. First, given a trained model $f_{\widehat{\BS{\theta}}}$, we utilize it as $f_{\widehat{\BS{\theta}}}(\M{x}_{i+j}),f_{\widehat{\BS{\theta}}}(\M{x}_{k+l})$ to estimate the missing ground-truth reference sources as $\tilde{\M{s}}_i,\tilde{\M{s}}_j,\tilde{\M{s}}_k,\tilde{\M{s}}_l$. Second, we apply the remixing strategy to generate unique mixtures such as $\tilde{\M{x}}_{j+k}=\tilde{\M{s}}_j+\tilde{\M{s}}_k$. Finally, we evaluate the trained model $f_{\widehat{\BS{\theta}}}$ on these artificial mixtures $\tilde{\M{x}}_{j+k}$. We repeat this procedure 100 times, which creates additional unique mixtures and uses them for evaluation. This can be viewed as applying dynamic mixing to a noisy version of the validation set, thus increasing its effective size and improving the reliability of the results.

Table~\ref{table:realm} compares different training setups according to ground-truth evaluation (GE), self-evaluation (SE) and mean opinion scores (MOS) on the validation sets of LibriMix and \realm{}, reporting means and standard deviations. We omit MixIT because it is not apparent how to select the correct two model outputs (out of four) when the reference sources are not available. The ground-truth evaluation and self-evaluation results on LibriMix are similar as expected. Therefore, we use self-evaluation (SE) as a surrogate for ground-truth evaluation (GE) on \realm{} and observe a considerable improvement ($+3.7$~dB) with MixCycle training on \realm{} over PIT-DM training on LibriMix.

As ground-truth evaluation is impossible on \realm{}, an informal listening test was conducted with 10 participants on 10 randomly-picked \realm{} validation mixtures to back up the self-evaluation results. The participants (4 female and 6 male), aged between 21 and 41 years, rated the separation results online while using headphones.
The participants were asked to give an overall quality score between 1-5 (higher is better), considering both the sound quality of the target source and the interference from the other source. We obtained mean opinion scores (MOS) as given in Table~\ref{table:realm}, which are consistent with the corresponding self-evaluation (SE) results. We also provide the audio samples\footnotemark[1] used in this test.

\begin{table}
\caption{Evaluation on the LibriMix test set in terms of SI-SNRi}
\label{table:results}
\setlength{\tabcolsep}{3pt}
\centering
\begin{tabular}{|l|c|r|r|r|r|}
\hline
\textbf{Method} & \textbf{Super.} & \textbf{5\%\,(dB)} & \textbf{100\%\,(dB)} & \textbf{Step\,(s)} & \textbf{Total\,(h)} \\
\hline
Conv-TasNet \cite{cosentino2020librimix}& Yes & -- & $14.7$ & -- & -- \\
IRM & Oracle & -- & $13.9{\pm}2.5$ & -- & -- \\ 
PIT & Yes & $7.1{\pm}5.0$ & $11.2{\pm}3.7$ & $0.26$ & $11.0$ \\
PIT-DM & Yes & $11.4{\pm}3.4$ & $11.8{\pm}3.2$ & $0.26$ & $11.9$ \\
\hline
MixIT (oracle) & No & $8.5{\pm}3.8$ & $9.9{\pm}3.4$ & $0.46$ & $27.6$ \\
MixIT & No & $6.0{\pm}3.5$ & $7.8{\pm}3.6$ & $0.46$ & $27.6$ \\
MixPIT (proposed) & No & $5.6{\pm}4.0$ & $7.1{\pm}3.8$ & $0.28$ & $4.2$ \\
MixCycle (proposed) & No & $\mathbf{11.2}{\pm}3.5$ & $\mathbf{11.4}{\pm}3.3$ & $0.32$ & $17.9$ \\
\hline
\end{tabular}
\end{table}

\begin{table}
\caption{Self-evaluation on the validation sets in terms of SI-SNRi}
\label{table:realm}
\setlength{\tabcolsep}{3pt}
\centering
\begin{tabular}{|l|l||r|r|r|r|}
\hline
\multicolumn{2}{ |c|| }{\textbf{Training Setup}} & \multicolumn{2}{ c| }{\textbf{LibriMix}} & \multicolumn{2}{ c| }{\textbf{\realm{}}} \\
\hline
\textbf{Method} & \textbf{Dataset} & \textbf{GE (dB)} & \textbf{SE (dB)} & \textbf{SE (dB)} & \textbf{MOS} \\
\hline
PIT-DM & LibriMix & $12.0{\pm}3.4$ & $12.0{\pm}3.9$ & $9.6{\pm}5.5$ & $2.9{\pm}1.2$ \\
MixPIT & LibriMix & $7.7{\pm}3.9$ & $7.0{\pm}3.6$ & $5.5{\pm}3.8$ & -- \\
MixCycle & LibriMix & $11.6{\pm}3.6$ & $11.9{\pm}4.0$ & $10.0{\pm}5.6$ & -- \\
\hline
MixCycle & \realm{} & $10.0{\pm}4.4$ & $10.0{\pm}4.7$ & $\mathbf{13.3}{\pm}4.5$ & $\mathbf{3.4}{\pm}1.2$ \\
\hline
\end{tabular}
\vspace{-0.1cm}
\end{table}

\vspace{-0.1cm}
\section{Conclusion}
We introduced unsupervised speech separation methods that avoid over-separation and narrow the performance gap between supervised and unsupervised training. We defer exploring mixtures with more than two sources and different source classes (these can pose a greater challenge in training the models) to future work. Also, we proposed a promising self-evaluation technique that we intend to investigate further.

\vspace{-0.1cm}
\section*{Acknowledgment}
We would like to thank Ali Taylan Cemgil, Cem Subakan and the anonymous reviewers for their insightful comments.

\bibliographystyle{IEEEtran}
\bibliography{refs}

\end{document}